\documentclass[%
 reprint,
 amsmath,amssymb,
 aps,
]{revtex4-2}

\usepackage{lipsum}
\usepackage{graphicx}% Include figure files
\usepackage{dcolumn}% Align table columns on decimal point
\usepackage{bm}% bold math
\usepackage{color}
\usepackage{comment}
\begin{document}
\preprint{APS/123-QED}
\title{Reduction of the  Twisted Bilayer Graphene Chiral Hamiltonian into a $2\times2$ matrix operator and physical origin of flat-bands at magic angles}
\author{Gerardo G. Naumis, Leonardo A. Navarro-Labastida, Enrique Aguilar-M\'endez, Abdiel Espinosa-Champo}
\date{February 2021}
\email{naumis@fisica.unam.mx}
\affiliation{%
Depto. de Sistemas Complejos, Instituto de F\'isica, \\ Universidad Nacional Aut\'onoma de M\'exico (UNAM)\\
Apdo. Postal 20-364, 01000, CDMX, M\'exico.
}%
\begin{abstract}
The chiral Hamiltonian for twisted graphene bilayers is written as a $2\times2$ matrix operator by a renormalization of the Hamiltonian that takes into account the particle-hole symmetry. {\color{blue}This results in an effective Hamiltonian written in terms of Pauli matrices}. The action of the proposed renormalization maps zero-modes flat-bands into ground states. On each graphene layer, modes near zero energy have an antibonding nature in a triangular lattice. This leads to a phase-frustration effect  associated with massive degeneration, and makes flat-bands modes similar to confined modes observed in other bipartite lattices. {\color{blue}At magic angles, it is shown that intralayer frustration is zero  while for other angles is proportional to the squared Fermi velocity.} Suprisingly, the proposed Hamiltonian renormalization suggests that flat-bands at magic angles are akin to floppy-mode bands in flexible crystals or glasses, making an unexpected connection between rigidity topological theory and twisted two-dimensional systems.
\end{abstract}

\maketitle
\textit{Introduction}. Superconducting states are difficult to reach as they require very strict laboratory parameters \cite{Nishijima2013}. High $T_c$ superconductors use cuprates which are well-ordered structures of atoms combined in three-dimensional arrangements \cite{Kudo2015,Keimer2015,Fu2020,Yamazaki2020,Nam2020,Tan2021}.  For these materials, the mechanism that counteracts the Colombian repulsion force between electrons is not exactly known, and for this reason, these unconventional states are referred to as strongly correlated \cite{Kyung2009,Capone2012}.
Recently, it has been discovered that  twisted bilayer graphene exhibits superconducting states at certain rotation angles \cite{Cao2018,Yankowitz2019} where the electron interactions are maximized \cite{Kerelsky2019}. This rotated graphene bilayer model generates a Moir\'e pattern as a function of the rotation angle, defining a Moir\'e Brillouin zone (mBZ) in reciprocal space. These special angles are called "magic" and were predicted as a possible consequence of flat-bands observed in previous theoretical work \cite{MacDonald2011}. In the work of Cao et.al. \cite{Cao2018},  a Mott insulating state appears in the middle of these superconducting phases. The study of the electronic properties of rotated graphene over graphene started before the discovery of superconductivity at magic angles.  In the work of J. Santos \cite{Santos2007} and A. Macdonald \cite{MacDonald2011}, a continuous Hamiltonian model was presented; however, due to the presence of an interlayer amplitude AA coupling, this model did not present chiral symmetry. In a recent work by G. Tarnopolsky et. al. \cite{Tarnpolsky2019}, a chiral continuum model was studied, and only the AB and BA inter-layer couplings are different from zero. Perhaps, so far, it is the simplest model that best captures the nature of magic angles; at these angles the dispersion energy becomes flat and has a recurrence behavior. At these magic angles the Fermi velocity also goes to zero.
Due to its chiral symmetry, the Hamiltonian of this model also  produces an intra-valley inversion symmetry \cite{Wang2020}, so the energy dispersion is inversion symmetric at all twist angles. Also, we can distinguish between different magic angles when zero modes occurs, and thus the inter-valley inversion classifies the topology of the twist angle. The zero-mode flat-band solutions have some resemblance to the ground state of a quantum Hall effect wave function on a torus \cite{Tarnpolsky2019,Popov2020}, and, therefore, the solution is of the harmonic oscillator type, where Landau levels arise \cite{Hejazi2019,Uri2020,Popov2020}. The mechanism that causes the appearance of these magic angles is still not known, however, many investigations suggest that it is a topological aspect of the band structure \cite{Qi2011,Zou2018,Xu2018,Fidrysiak2018,You2019,Khalaf2020,Vogl2020}. The aim of this work its to clarify the physical behavior of this model and to develop an effective equation for the Hamiltonian matrix in order to study all other non flat band states. 
In particular, here we renormalize the Hamiltonian to take into account the particle-hole symmetry which results in a folding of the spectrum around zero-energy. Then we discuss the physical picture that arises from the renormalization

\textit{Twisted Bilayer Graphene Effective Model}. Tarnopolsky et. al. derived a chiral Hamiltonian for electrons-holes in twisted bilayer graphene. It captures the "true magic" of the magic angle physics \cite{Tarnpolsky2019},
\begin{equation}\label{eq:Hstart}
\begin{split}
\mathcal{H}
&=\begin{pmatrix} 
0 & D^{\ast}(-r)\\
 D(r) & 0
  \end{pmatrix}  \\
\end{split} 
\end{equation}
where the zero-mode operator is, 
\begin{equation}
\begin{split}
D(r)&=\begin{pmatrix} 
-i\Bar{\partial} & \alpha U(r)\\
  \alpha U(-r) & -i\Bar{\partial} 
  \end{pmatrix}  \\
\end{split} 
\end{equation}
and,
\begin{equation}
\begin{split}
D^{*}(-r)&=\begin{pmatrix} 
-i\partial & \alpha U^{*}(-r)\\
  \alpha U^{*}(r) & -i\partial 
  \end{pmatrix}  \\
\end{split} 
\end{equation}

with $\Bar{\partial}=\partial_x+i\partial_y$, $\partial=\partial_x-i\partial_y$. The potential is,
\begin{equation}
    U(\bm{r})=e^{-i\bm{q}_1\cdot \bm{r}}+e^{i\phi}e^{-i\bm{q}_2\cdot \bm{r}}+e^{-i\phi}e^{-i\bm{q}_3\cdot \boldsymbol{r}}
\end{equation}

\begin{figure}[h!]
    \includegraphics[width=9cm]{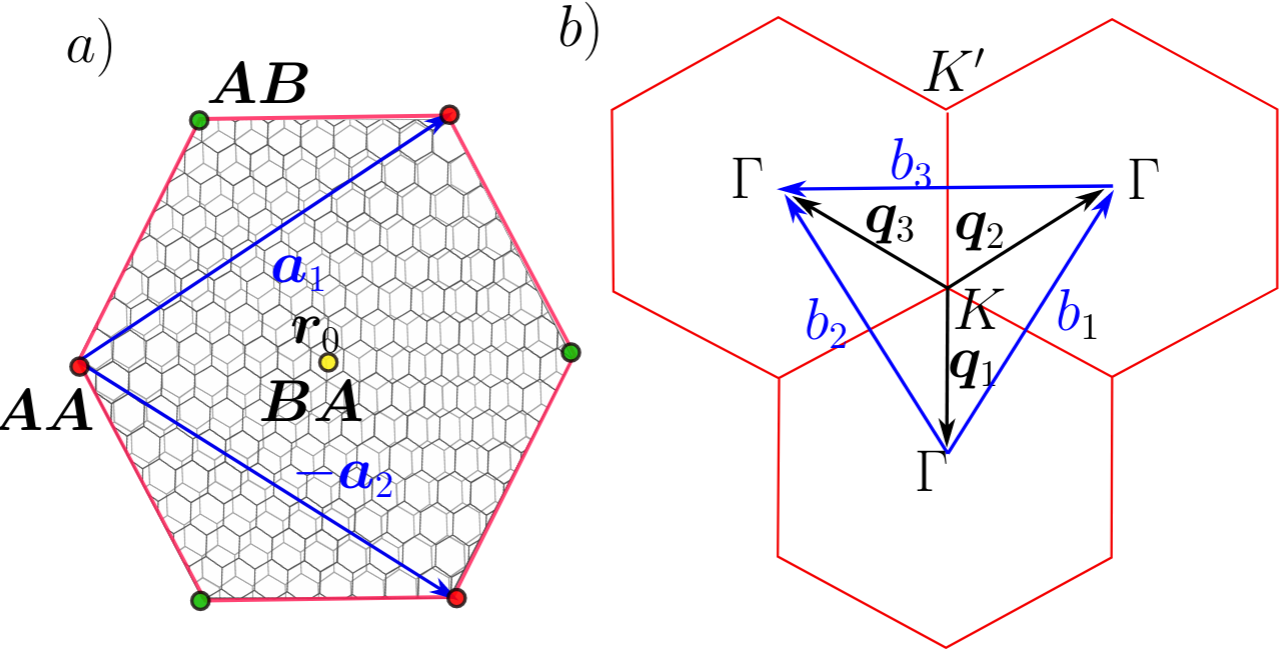}
    \caption{ a) Real space Moir\'e unit cell, $\bm{a_{1,2}}$ are two Moiré lattice vectors. Point $\bm{r}_0=(\bm{a}_1-\bm{a}_2)/3$ is the BA stacking point where all components of the wave function vanish at magic $\alpha$. b) mBZ in reciprocal space, $\bm{b_{1,2}}$ are the base vectors.}
    \label{fig:zonas}
\end{figure}

For this Hamiltonian, the parameters are $\phi=\frac{2\pi}{3}$ and $\bm{q}_1=k_\theta(0,-1)$, $\bm{q}_2=k_\theta(\frac{\sqrt{3}}{2},\frac{1}{2})$ and $\bm{q}_3=k_\theta(-\frac{\sqrt{3}}{2},\frac{1}{2})$, the Moir\'e modulation vector is $k_\theta=2k_D\sin{\frac{\theta}{2}}$ with $k_D=\frac{4\pi}{3a_0}$ is the magnitude of the Dirac wave vector and $a_0$ is the lattice constant of monolayer graphene. The physics of this
model is captured by the parameter $\alpha$, defined as $\alpha=\frac{w_1}{v_0k_\theta}$. Here $w_1$ is the interlayer coupling of stacking AB and BA, take the value  $w_1=110$meV and $v_0$ is the Fermi velocity, with value $v_0=\frac{19.81eV}{2k_D}$.At magic angles $\alpha=0.586, 2.221, 3.751, 5.276, 6.795, 8.313, 9.829, 11.345, ...$, flat-bands appear. Magic $\alpha$'s follow a remarkable $3/2$ quantization rule \cite{Tarnpolsky2019} for $\alpha>0.586$ . 

This Hamiltonian is difficult to tackle and in fact most of the studies have been restricted to the zero-mode operator solutions at energy zero \cite{Wang2020,Popov2020}. Here, instead of solving the Schr$\ddot{o}$dinger equation with $\mathcal{H}$
we first propose to reduce the dimensionality of the problem. Starting with the Schr$\ddot{o}$dinger equation $\mathcal{H} \Phi=E \Phi$, where $\Phi(r)=\begin{pmatrix} 
\psi_1(r) ,
\psi_2(r),
\chi_1(r),
\chi_2(r)
\end{pmatrix}^T$ are the four components of the twisted graphene bilayer, and the index $1,2$ represent each graphene layer,  we consider the squared Hamiltonian $\mathcal{H}^{2}$,
\begin{equation}
\begin{split}
\mathcal{H}^{2}&=\begin{pmatrix} D^{\ast}(-r)D(r) & 0\\
 0 & D(r)D^{\ast}(-r)
  \end{pmatrix} 
  \end{split} 
  \end{equation}
This transforms the off-diagonal blocks into zeros. We can understand such transformation as a removal of the particle-hole symmetry that is an anti-unitary anti-commuting symmetry. Therefore, we obtain a decoupled equation  $\mathcal{H}^{2}\Phi(r)=E^{2}\Phi(r)$, where
the eigenvalues are the squares of the original energies. Therefore, the states at $E=0$ are ground states of $\mathcal{H}^{2}\Phi(r)$. For arbitray $\alpha$ there are always two zero-mode solutions in the $K$ and $K'$ points \cite{Tarnpolsky2019}, the signature of a magic angle in $\mathcal{H}^{2}$ is that the Fermi velocity at $\bm{k}=K,K'$ points of the Moir\'e  Brillouin  (mBrillouin) zone  approaches zero and a massive degenerate ground state. We now define a $2\times 2$ effective Hamiltonian $H^{2}=D^{\ast}(-r)D(r)$. As detailed in the supplementary material, the resulting  effective Hamiltonian is,

\begin{equation}\label{eq:H2}
\begin{split}
H^{2}&=\begin{pmatrix} -\nabla^{2}+\alpha^{2} |U(\bm{-r})|^{2} & \alpha A^{\dagger}(\bm{r})\\
 \alpha A(\bm{r}) & -\nabla^{2}+\alpha^{2} |U(\bm{r})|^{2}
  \end{pmatrix} 
  \end{split} 
  \end{equation}
The norm of the potential is,
\begin{equation}\label{eq:Potdef}
|U(\bm{r})|^{2}=3+2 \cos(\bm{b}_1\cdot \bm{r}-\phi )+2\cos(\bm{b}_2\cdot \bm{r}+\phi )+2\cos(\bm{b}_3\cdot \bm{r}+2\phi)    \end{equation}
where $\bm{b}_{1,2}=\bm{q}_{2,3}-\bm{q}_{1}$ are the mBrillouin zone Moir\'e vectors and $\bm{b}_{3}=\bm{q}_{3}-\bm{q}_2$. 
The off-diagonal terms are,
\begin{equation}\label{eq:Adef}
 \begin{split}
A(\bm{r}) & =-i\sum_{\mu=1}^{3}e^{i\bm{q}_{\mu}\cdot\bm{r}}(2\bm{\hat{q}}_{\mu}^{\perp}\cdot \bm{\nabla}-k_{\theta})\\
  \end{split} 
\end{equation}
and,
\begin{equation}\label{eq:Adef2}
 \begin{split}
A^{\dagger}(\bm{r}) & =-i\sum_{\mu=1}^{3}e^{-i\bm{q}_{\mu}\cdot\bm{r}}(2\bm{\hat{q}}_{\mu}^{\perp}\cdot \bm{\nabla}+k_{\theta})\\
  \end{split} 
\end{equation}
where  $\bm{\nabla}^{\dagger}=-\bm{\nabla}$ with $\bm{\nabla}=(\partial_x,\partial_y)$ and $\mu=1,2,3$. This is an essential point as eigenvalues must be reals (notice that $-A^{\dagger}(\bm{-r})=A(\bm{r})$). Also, $\bm{\hat{q}}_{\mu}^{\perp}$ is a set of unitary vectors perpendicular to the set $\bm{q}_{\mu}$ (see supplementary).  {\color{blue}The term $\bm{\hat{q}}_{\mu}^{\perp} \cdot \bm{\nabla}$ is the directional gradient along the triangle defined by the Moiré vectors and has  an interpretation in terms of frustration (see supplementary).}

Any eigenfunction of the original Hamiltonian is an eigenfunction of $\mathcal{H}^{2}$, therefore, the spinor $\Psi(\bm{r})=(\psi_1(\bm{r}),\psi_2(\bm{r}))$ is a solution of $H^{2}$. In a similar way, we can proceed to use $D(r)D^{\ast}(-r)$ with solutions $(\chi_1(\bm{r}),\chi_2(\bm{r}))$. 
These solutions are easily obtained from $(\psi_1(\bm{r}),\psi_2(\bm{r}))$ using symmetry operations, and thus here we only study $H^{2}$.
$\mathcal{H}^{2}$ eigenfunctions are made from a superposition of pseudo-spin polarized states of $\mathcal{H}$.
%Using the operator defined in (\ref{eq:H2}) and the SchrÃ¶dinger equation $H^{2}\Psi(\bm{r})=E^{2}\Psi(\bm{r})$, it is possible to realize that by making the replacement $\bm{r}\rightarrow -\bm{r}$, we have $A^{\dagger}(\bm{r})=-A(-\bm{r})$  resulting in $\psi_2(\bm{r})=i\psi_1(\bm{-r})$. A similar situation holds for the lower part of the bispinor. 
%Thus, the solution has the form $\Phi(\bm{r})=\begin{pmatrix}
%\psi_1(\bm{r}),i\psi_1(\bm{-r}),\chi_1(\bm{r}),i\chi_1(\bm{-r})
%\end{pmatrix}^T$. 

\begin{figure}[h!]
    \includegraphics[scale=0.25]{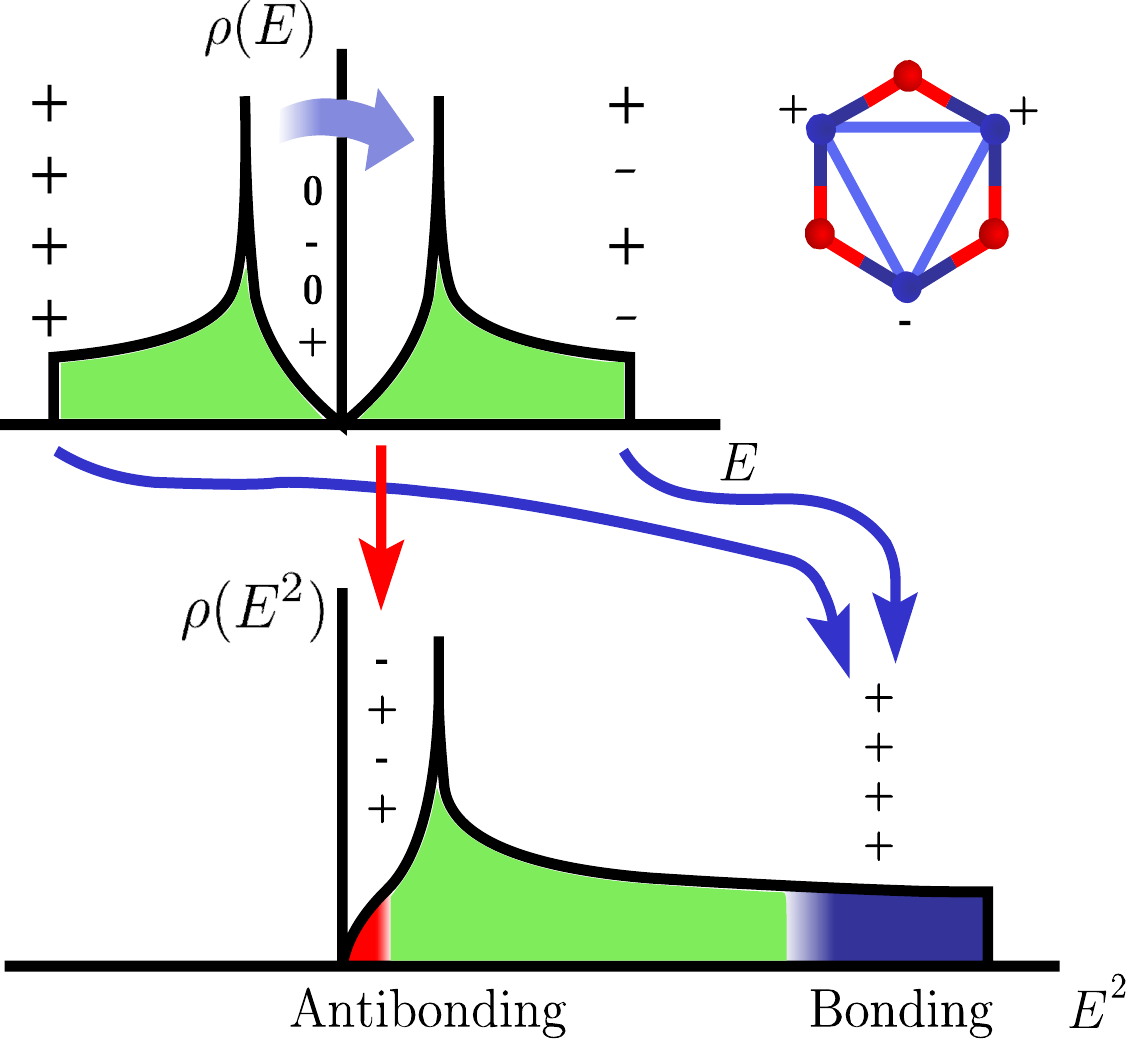}
    \caption{ {\color{blue} For each non-interacting monolayer graphene case, we present the density of states $\rho(E)$ and of $\rho(E^{2})$ corresponding to the squared Hamiltonian, equivalent to a folding of the spectrum. The signs are a sketch of the relative signs and amplitudes on bipartite sublattices A and B (shown in red and blue). Arrows indicate how the states are mapped under such transformation. Zero modes and close-by states are in the antibonding limit on a triangular lattice (indicated by the triangle inside the hexagon). The frustration is seen here in the bond that joins the $+$  and $+$ sites in the triangle (in reality, the minimal phase difference is $\phi$). For the twisted bilayer with interlayer interaction, $\mathcal{H}^{2}$ produces a similar DOS folding and relative signs for the superlattice.}}
    \label{fig:Renormalization}
\end{figure}

\textit{Physical interpretation of the effective model and flat-bands} The renormalization into a $2\times 2$ matrix not only simplifies
the mathematics but has a profound physical meaning. What we achieved is the decoupling of the $A$ and $B$ bipartite lattices for each graphene layer. In graphene, such squared Hamiltonian is equivalent to renormalize the honeycomb lattice into a triangular lattice with renormalized interactions and a self-energy \cite{Naumis2007,Barrios_Vargas_2011}, here Eq. (\ref{eq:H2}) shows that this is also the case. {\color{blue}As sketched in Fig. \ref{fig:Renormalization} for monolayer graphene, the transformation of $E$ to $E^{2}$ produces a fold in the spectrum such that bonding states with the lowest energy are mapped into higher energies}. In other words, the transformation deletes the alternating sign of one of the bipartite sublattices as explained in Fig. \ref{fig:Renormalization}.

As the zero-modes are ground states of $H^{2}$, we end up having a clear picture of nearby states around zero-energy modes, they correspond to the antibonding limit in {\color{blue}two coupled triangular lattices}. Anti-bonding states in non-bipartite lattices are frustrated as they cannot achieve  a phase difference of $\pi$ between sites as odd-rings are present \cite{Cohen1981,Cohen1983}. For disordered systems, states are localized in regions of lower frustration and a kind of Lifshitz tail appears \cite{Cohen1983}. Moreover, frustration is always associated with massive degeneration \cite{Cohen1981,Cohen1983}; it leads to Van Hove singularites or if possible, in a condensation of confined states. These confined states appear in chiral models of the Penrose lattice \cite{Kohmoto1986,Arai1988,Naumis1994}, where they form beautiful fractal patterns \cite{Arai1988}, in random binary alloys \cite{Eggarter1972,Naumis2002} and in graphene with defects \cite{Naumis2007,Barrios2013}. Strictly confined states are degenerate. However, there is a basis in which the amplitude in one of the bipartite sublattices is zero while in the other, the sum of all neighbors amplitudes  is always zero for any site \cite{Eggarter1972,Arai1988}. 

%This last equation is simplified if we use, $\zeta=u_1$ and $\eta=u_2-u_3$,
%\begin{equation}\label{eq:U2inpseta}
%|U(\zeta,\eta)|^{2}=3+2\cos{\zeta}+4\cos{\frac{\zeta}{2}}\cos{\frac{\eta}{2}}
%  \end{equation}
%where $0\leq\eta\leq4\pi$ and $0\leq\zeta\leq4\pi$. $\zeta$ and $\eta$ are coordinates %that indicate the departure from the medians of a triangle

{\color{blue}To understand how zero modes are related with confined states and  frustration as in Fig. \ref{fig:Renormalization}, let us made the following remarks. From the Hamiltonian Eq. (\ref{eq:Hstart}) we confirm  that for $E=0$ there are always solutions of the form $\Phi_{\boldsymbol{k}}^{A}(\boldsymbol{r})=(\psi_{\boldsymbol{k},1}^{A}(\boldsymbol{r}),\psi_{\boldsymbol{k},2}^{A}(\boldsymbol{r}),0,0)$ and $\Phi_{\boldsymbol{k}}^{B}(\boldsymbol{r})=(0,0,\chi_{\boldsymbol{k},1}^{B}(\boldsymbol{r}),\chi_{\boldsymbol{k},2}^{B}(\boldsymbol{r}))$, where the labels $A$ and $B$ are used to denote
zero amplitude in the opposite bipartite sublattice. We remark that linear combinations, 
\begin{equation}
    \Phi_{\boldsymbol{k}}(\boldsymbol{r})=\frac{1}{\sqrt{2}}\left(\Phi_{\boldsymbol{k}}^{A}(\boldsymbol{r})+e^{i\gamma}\Phi_{\boldsymbol{k}}^{B}(\boldsymbol{r})\right)
\end{equation}
with $\gamma$ a phase, result in a different basis which do not show zeros in one sublattice, as for example with the symmetrized/antisymmetrized cases $\gamma=0,\pi$. As the potential does not brake the $C_3$ symmetry, the states $\boldsymbol{k}=\boldsymbol{K},\boldsymbol{K}'$ are always a $E=0$ solution for any $\alpha$. As a conclusion, for $\alpha$ not a magical angle there are four linearly independent wavefunctions, as confirmed from a Wronskian analisis \cite{Popov2020} and therefore, at any angle there are "confined states" in the sense of Fig. \ref{fig:Renormalization}. At magic angles, the
Wronskian of the solutions is zero and there are  $E=0$ solutions at any $\boldsymbol{k}$, resulting in the flat band. Still, $\Phi_{\boldsymbol{k}}^{A}(\boldsymbol{r})$ and $\Phi_{\boldsymbol{k}}^{B}(\boldsymbol{r})$ are solutions meaning that now all states are "confined". Any linear combination using different sets of $\boldsymbol{k}$ is a solution. As explained in the supplementary, this is similar to the Van-Hove singularity in monolayer graphene, where dimers are desconnected from the lattice and thus is a kind of highly-degenerated, confined state.

The anti-bonding or bonding nature and therefore frustration is obtained from all bonds energy contribution, the latter one obtained from the product of the wavefunction in a site with  the conjugated wavefunction of a neighboring site \cite{Naumis2002}. For the present analysis, this requires to take into account three factors: 1) the system has two layers, 2) we are dealing with a low-energy continuous version of the TBH and 3) the system has a superlattice. Concerning point 1), we look at the intralayer frustration  to see how the interlayer interaction tunes such contribution. Points 2) and 3) are more delicate as we need to understand that $k$ is a moment that departs from $K$ and $K'$. In a two-layer bipartite continuous lattice, for a given state $\boldsymbol{k}$ such procedure is equivalent to consider bonds joining $A$ and $B$ sublattices sites . Although this can be made using any basis, it is easier to use in a symmetrized one $\gamma=0$. The frustration can be measured from the function (see supplementary),
\begin{equation}
    g_{\boldsymbol{k}}(\boldsymbol{r})= \psi_{\boldsymbol{k},1}(\boldsymbol{r}) \chi_{\boldsymbol{k},1}^{*}(\boldsymbol{r})+ \psi_{\boldsymbol{k},2}(\boldsymbol{r}) \chi_{\boldsymbol{k},2}^{*}(\boldsymbol{r}) 
    \label{eq:Fk}
\end{equation}
 and $\nabla g_{\boldsymbol{k}}(\boldsymbol{r})$. As we are interested in states near $E=0$,  we set $\boldsymbol{k}=\boldsymbol{K}$ in Eq. (\ref{eq:Fk}). Using the symmetry of the problem, we can show that $\chi_{\boldsymbol{K}}^{*}(\boldsymbol{r})=\psi_{\boldsymbol{K}}(-\boldsymbol{r})$ and,
\begin{equation}
    g_{\boldsymbol{K}}(\boldsymbol{r})= \psi_{\boldsymbol{K},1}(\boldsymbol{r}) \psi_{\boldsymbol{K},1}(-\boldsymbol{r})+ \psi_{\boldsymbol{K},2}(\boldsymbol{r}) \psi_{\boldsymbol{K},2}(-\boldsymbol{r})\sim v_F(\alpha) 
    \label{eq:Falpha}
\end{equation}
where this last step is obtained from the fact that $g_{\boldsymbol{K}}$ turns out to be an invariant (see supplementary) which can be identified with the Fermi velocity at a given angle \cite{Tarnpolsky2019}, here denoted by $v_{F}(\alpha)$. For any angle $\nabla g_{\boldsymbol{K}}(\boldsymbol{r})=0$ indicating that $\boldsymbol{K}=0$ is minimally frustrated. However, at magic angles $v_F(\alpha)=0$ and therefore also  $g_{\boldsymbol{K}}=0$, i.e., frustration is zero, making a flat band by pushing all states towards $E=0$. This leads to a special condition for the wave function Fourier components. The explicit form of the wavefunction is,
\begin{equation}
    \begin{split} \psi_{\boldsymbol{k}}(\boldsymbol{r})=&\sum_{m,n}\begin{pmatrix} 
a_{mn}\\
b_{mn}e^{i\boldsymbol{q}_1\cdot\boldsymbol{r}} 
  \end{pmatrix} \\
\end{split} e^{i(\boldsymbol{K}_{mn}+\boldsymbol{k})\cdot\boldsymbol{r}} 
\label{eq:wave}
\end{equation}
 where $\boldsymbol{K}_{mn}=m\boldsymbol{b}_1+n\boldsymbol{b}_2$. $a_{m,n}$ and $b_{m,n}$ are the Fourier coefficients. At the $BA$ stacking point $\boldsymbol{r}_0=(\boldsymbol{a}_1-\boldsymbol{a}_2)/3$, and for any $\alpha$ and due to symmetry reasons $\psi_{\boldsymbol{K},2}(\boldsymbol{r}_0)\psi_{\boldsymbol{K},2}(-\boldsymbol{r}_0)=0$ , while for magic $\alpha$ we have that \cite{Tarnpolsky2019}  $\psi_{\boldsymbol{K},1}(\boldsymbol{r}_0)=0$. As shown in the supplementary, when translated into Eq. (\ref{eq:wave}) results in, 
\begin{equation}
 \begin{split}
    \sum_{m,s} \left(a_{m,3s}+a_{m+1,3s}e^{i\phi}+a_{m,3s+1}e^{-i\phi} \right)=0
\end{split} 
\label{eq:condition}
\end{equation}
and the same equation holds for $b_{mn}$, a fact to be expected as the Wronskian is zero \cite{Popov2020}. The previous equation shows a precise tuning of components and hints how states develop a sharp peak at the AA stacking point as $\alpha \rightarrow \infty$ (see supplementary).

Quite surprisingly, our renormalization  is akin to a phonon problem (see supplementary). Thus, flat-band states seem to have a  resemblance to floppy modes in an equivalent rigidity-phonon problem, i.e., the flat-band can be interpreted as a massive zero frequency vibrational band, since $E^{2}$ is analogous to a frequency \cite{Kane2014}. These floppy modes are well known in the Phillips rigidity theory of glasses \cite{Phillips,Huerta2002,Flores2010} and are reminiscent of the protected electronic boundary modes that occur in the quantum Hall effect and in topological insulators\cite{Kane2014}.\\
} 
\textit{Effective fields and Hamiltonian in triangular coordinates} It is useful to write the renormalized twisted graphene bilayer Hamiltonian in terms of the Pauli matrix vector $\hat{\bm{\sigma}}$ as follows.

\begin{equation}
    H^{2}=h_0(\bm{r})\sigma_0+\bm{h}(\bm{r})\cdot \hat{\bm{\sigma}}
\end{equation}
where the vector $\bm{h}(\bm{r})$ is,
\begin{equation}
    \bm{h}(\bm{r})=(\alpha h_x(\bm{r}), \alpha h_y(\bm{r}),\alpha^{2}h_z(\bm{r}))
\end{equation}
with,
\begin{equation}
\begin{split}
h_x(\bm{r})&=
-\sum_{\mu}[k_{\theta}\sin{(\bm{q}_{\mu}\cdot\bm{r})}+2i\cos{(\bm{q}_{\mu}\cdot\bm{r})}\bm{\hat{q}}_{\mu}^{\perp}\cdot \bm{\nabla}] \\
h_y(\bm{r})&=\sum_{\mu}[k_{\theta}\cos{(\bm{q}_{\mu}\cdot\bm{r})}-2i\sin{(\bm{q}_{\mu}\cdot\bm{r})}\bm{\hat{q}}_{\mu}^{\perp}\cdot \bm{\nabla}]\\
h_z(\bm{r}) &= \frac{ |U(\bm{-r})|^{2}-|U(\bm{r})|^{2}}{2}\end{split} 
\end{equation}

The operator in front of the identity $\sigma_0$ is,
\begin{equation}
    h_0(\bm{r})=-\nabla^{2}+ \alpha^{2} \bar{V}(\bm{r})
  \end{equation}

\begin{equation}
\begin{split}
\bar{V}(\bm{r})=\frac{|U(\bm{r})|^{2}+|U(\bm{-r})|^{2}}{2}
\end{split}
\end{equation}

We observe that $h_0(\bm{r})$ corresponds to  a Hamiltonian with an average potential. The Hamiltonian structure is akin to the one found in Ref. \cite{Guinea2012}. Observe that the term $\bm{h}(\bm{r})$ contains all the topological properties of the operator. Let us further simplify this Hamiltonian. Define $\psi_{\pm}(\bm{r})=\psi_1(\bm{r})\pm \psi_2(\bm{r})$. The Schr\"odinger equation is transformed into $
H_{\text{eff}}(\psi_{+}(\bm{r}),\psi_{-}(\bm{r}))^{T}=E^{2}(\psi_{+}(\bm{r}),\psi_{-}(\bm{r}))^{T}$. The stated effective Hamiltonian is,
\begin{equation}
\begin{split}
H_{\text{eff}}&=\begin{pmatrix} 
-\nabla^{2}+V_{\text{eff}}(\bm{r}) & A_{\text{eff}}^{\dagger}(\bm{r})\\
   A_{\text{eff}}(\bm{r}) &  -\nabla^{2}+V_{\text{eff}}(\bm{r}) 
  \end{pmatrix}  \\
\end{split} 
\end{equation}

Here we defined the effective potentials as,
\begin{equation}
    V_{\text{eff}}(\bm{r})=\alpha^{2} \bar{V}(\bm{r})+\alpha h_x(\bm{r})
\end{equation}
and,
\begin{equation}
    A_{\text{eff}}(\bm{r})=\alpha^{2} h_z(\bm{r})+i\alpha h_y(\bm{r})
\end{equation}

In the supplementary section, we show how this system can be simplified  using triangular coordinates to map the problem into a friendly rectangular domain.

\textit{Conclusions} We showed that the chiral Hamiltonian for twisted graphene bilayers can be written into a $2\times2$ matrix operator. The action of the proposed renormalization maps the zero-mode region into the ground state. Modes next to zero energy have an antibonding nature in a triangular lattice and at zero energy are similar to confined modes observed in many other bipartite systems \cite{Eggarter1972,Naumis1994,Naumis2002}. A surprising result is that our renormalization suggests that flat-bands are somehow analogous to floppy modes in  rigidity phonon models  \cite{Kane2014,Naumis2002,Dasgupta}.

We thank UNAM-DGAPA project IN102620 and CONACyT project 1564464. 

%\bibliographystyle{plain}
%\bibliography{MagicAngleReferences.bib}

\begin{thebibliography}{42}%
	\makeatletter
	\providecommand \@ifxundefined [1]{%
		\@ifx{#1\undefined}
	}%
	\providecommand \@ifnum [1]{%
		\ifnum #1\expandafter \@firstoftwo
		\else \expandafter \@secondoftwo
		\fi
	}%
	\providecommand \@ifx [1]{%
		\ifx #1\expandafter \@firstoftwo
		\else \expandafter \@secondoftwo
		\fi
	}%
	\providecommand \natexlab [1]{#1}%
	\providecommand \enquote  [1]{``#1''}%
	\providecommand \bibnamefont  [1]{#1}%
	\providecommand \bibfnamefont [1]{#1}%
	\providecommand \citenamefont [1]{#1}%
	\providecommand \href@noop [0]{\@secondoftwo}%
	\providecommand \href [0]{\begingroup \@sanitize@url \@href}%
	\providecommand \@href[1]{\@@startlink{#1}\@@href}%
	\providecommand \@@href[1]{\endgroup#1\@@endlink}%
	\providecommand \@sanitize@url [0]{\catcode `\\12\catcode `\$12\catcode
		`\&12\catcode `\#12\catcode `\^12\catcode `\_12\catcode `\%12\relax}%
	\providecommand \@@startlink[1]{}%
	\providecommand \@@endlink[0]{}%
	\providecommand \url  [0]{\begingroup\@sanitize@url \@url }%
	\providecommand \@url [1]{\endgroup\@href {#1}{\urlprefix }}%
	\providecommand \urlprefix  [0]{URL }%
	\providecommand \Eprint [0]{\href }%
	\providecommand \doibase [0]{https://doi.org/}%
	\providecommand \selectlanguage [0]{\@gobble}%
	\providecommand \bibinfo  [0]{\@secondoftwo}%
	\providecommand \bibfield  [0]{\@secondoftwo}%
	\providecommand \translation [1]{[#1]}%
	\providecommand \BibitemOpen [0]{}%
	\providecommand \bibitemStop [0]{}%
	\providecommand \bibitemNoStop [0]{.\EOS\space}%
	\providecommand \EOS [0]{\spacefactor3000\relax}%
	\providecommand \BibitemShut  [1]{\csname bibitem#1\endcsname}%
	\let\auto@bib@innerbib\@empty
	%</preamble>
	\bibitem [{\citenamefont {{Nishijima}}\ \emph {et~al.}(2013)\citenamefont
		{{Nishijima}}, \citenamefont {{Eckroad}}, \citenamefont {{Marian}},
		\citenamefont {{Choi}}, \citenamefont {{Kim}}, \citenamefont {{Terai}},
		\citenamefont {{Deng}}, \citenamefont {{Zheng}}, \citenamefont {{Wang}},
		\citenamefont {{Umemoto}}, \citenamefont {{Du}}, \citenamefont {{Febvre}},
		\citenamefont {{Keenan}}, \citenamefont {{Mukhanov}}, \citenamefont
		{{Cooley}}, \citenamefont {{Foley}}, \citenamefont {{Hassenzahl}},\ and\
		\citenamefont {{Izumi}}}]{Nishijima2013}%
	\BibitemOpen
	\bibfield  {author} {\bibinfo {author} {\bibfnamefont {S.}~\bibnamefont
			{{Nishijima}}}, \bibinfo {author} {\bibfnamefont {S.}~\bibnamefont
			{{Eckroad}}}, \bibinfo {author} {\bibfnamefont {A.}~\bibnamefont {{Marian}}},
		\bibinfo {author} {\bibfnamefont {K.}~\bibnamefont {{Choi}}}, \bibinfo
		{author} {\bibfnamefont {W.~S.}\ \bibnamefont {{Kim}}}, \bibinfo {author}
		{\bibfnamefont {M.}~\bibnamefont {{Terai}}}, \bibinfo {author} {\bibfnamefont
			{Z.}~\bibnamefont {{Deng}}}, \bibinfo {author} {\bibfnamefont
			{J.}~\bibnamefont {{Zheng}}}, \bibinfo {author} {\bibfnamefont
			{J.}~\bibnamefont {{Wang}}}, \bibinfo {author} {\bibfnamefont
			{K.}~\bibnamefont {{Umemoto}}}, \bibinfo {author} {\bibfnamefont
			{J.}~\bibnamefont {{Du}}}, \bibinfo {author} {\bibfnamefont {P.}~\bibnamefont
			{{Febvre}}}, \bibinfo {author} {\bibfnamefont {S.}~\bibnamefont {{Keenan}}},
		\bibinfo {author} {\bibfnamefont {O.}~\bibnamefont {{Mukhanov}}}, \bibinfo
		{author} {\bibfnamefont {L.~D.}\ \bibnamefont {{Cooley}}}, \bibinfo {author}
		{\bibfnamefont {C.~P.}\ \bibnamefont {{Foley}}}, \bibinfo {author}
		{\bibfnamefont {W.~V.}\ \bibnamefont {{Hassenzahl}}},\ and\ \bibinfo {author}
		{\bibfnamefont {M.}~\bibnamefont {{Izumi}}},\ }\bibfield  {title} {\bibinfo
		{title} {{Superconductivity and the environment: a Roadmap}},\ }\href
	{https://doi.org/10.1088/0953-2048/26/11/113001} {\bibfield  {journal}
		{\bibinfo  {journal} {Superconductor Science Technology}\ }\textbf {\bibinfo
			{volume} {26}},\ \bibinfo {eid} {113001} (\bibinfo {year}
		{2013})}\BibitemShut {NoStop}%
	\bibitem [{\citenamefont {Kudo}\ \emph {et~al.}(2015)\citenamefont {Kudo},
		\citenamefont {Yoshida}, \citenamefont {Ideta}, \citenamefont {Takashima},
		\citenamefont {Anzai}, \citenamefont {Fujita}, \citenamefont {Nakashima},
		\citenamefont {Ino}, \citenamefont {Arita}, \citenamefont {Namatame},
		\citenamefont {Taniguchi}, \citenamefont {Kojima}, \citenamefont {Uchida},\
		and\ \citenamefont {Fujimori}}]{Kudo2015}%
	\BibitemOpen
	\bibfield  {author} {\bibinfo {author} {\bibfnamefont {S.}~\bibnamefont
			{Kudo}}, \bibinfo {author} {\bibfnamefont {T.}~\bibnamefont {Yoshida}},
		\bibinfo {author} {\bibfnamefont {S.}~\bibnamefont {Ideta}}, \bibinfo
		{author} {\bibfnamefont {K.}~\bibnamefont {Takashima}}, \bibinfo {author}
		{\bibfnamefont {H.}~\bibnamefont {Anzai}}, \bibinfo {author} {\bibfnamefont
			{T.}~\bibnamefont {Fujita}}, \bibinfo {author} {\bibfnamefont
			{Y.}~\bibnamefont {Nakashima}}, \bibinfo {author} {\bibfnamefont
			{A.}~\bibnamefont {Ino}}, \bibinfo {author} {\bibfnamefont {M.}~\bibnamefont
			{Arita}}, \bibinfo {author} {\bibfnamefont {H.}~\bibnamefont {Namatame}},
		\bibinfo {author} {\bibfnamefont {M.}~\bibnamefont {Taniguchi}}, \bibinfo
		{author} {\bibfnamefont {K.~M.}\ \bibnamefont {Kojima}}, \bibinfo {author}
		{\bibfnamefont {S.}~\bibnamefont {Uchida}},\ and\ \bibinfo {author}
		{\bibfnamefont {A.}~\bibnamefont {Fujimori}},\ }\bibfield  {title} {\bibinfo
		{title} {Temperature evolution of correlation strength in the superconducting
			state of high-${T}_{c}$ cuprates},\ }\href
	{https://doi.org/10.1103/PhysRevB.92.195135} {\bibfield  {journal} {\bibinfo
			{journal} {Phys. Rev. B}\ }\textbf {\bibinfo {volume} {92}},\ \bibinfo
		{pages} {195135} (\bibinfo {year} {2015})}\BibitemShut {NoStop}%
	\bibitem [{\citenamefont {Keimer}\ \emph {et~al.}(2015)\citenamefont {Keimer},
		\citenamefont {Kivelson},\ and\ \citenamefont {Norman}}]{Keimer2015}%
	\BibitemOpen
	\bibfield  {author} {\bibinfo {author} {\bibfnamefont {B.}~\bibnamefont
			{Keimer}}, \bibinfo {author} {\bibfnamefont {S.}~\bibnamefont {Kivelson}},\
		and\ \bibinfo {author} {\bibfnamefont {M.~e.~a.}\ \bibnamefont {Norman}},\
	}\bibfield  {title} {\bibinfo {title} {From quantum matter to
			high-temperature superconductivity in copper oxides},\ }\href
	{https://doi.org/10.1038/nature14165} {\bibfield  {journal} {\bibinfo
			{journal} {Nature}\ }\textbf {\bibinfo {volume} {518}},\ \bibinfo {pages}
		{179–186} (\bibinfo {year} {2015})}\BibitemShut {NoStop}%
	\bibitem [{\citenamefont {{Wang}}\ \emph
		{et~al.}(2020{\natexlab{a}})\citenamefont {{Wang}}, \citenamefont {{Yuan}},\
		and\ \citenamefont {{Fu}}}]{Fu2020}%
	\BibitemOpen
	\bibfield  {author} {\bibinfo {author} {\bibfnamefont {T.}~\bibnamefont
			{{Wang}}}, \bibinfo {author} {\bibfnamefont {N.~F.~Q.}\ \bibnamefont
			{{Yuan}}},\ and\ \bibinfo {author} {\bibfnamefont {L.}~\bibnamefont {{Fu}}},\
	}\bibfield  {title} {\bibinfo {title} {{Moir{\'e} surface states and enhanced
				superconductivity in topological insulators}},\ }\href@noop {} {\bibfield
		{journal} {\bibinfo  {journal} {arXiv e-prints}\ ,\ \bibinfo {eid}
			{arXiv:2010.09753}} (\bibinfo {year} {2020}{\natexlab{a}})},\ \Eprint
	{https://arxiv.org/abs/2010.09753} {arXiv:2010.09753 [cond-mat.supr-con]}
	\BibitemShut {NoStop}%
	\bibitem [{\citenamefont {{Yamazaki}}\ \emph {et~al.}(2020)\citenamefont
		{{Yamazaki}}, \citenamefont {{Ochi}}, \citenamefont {{Ogura}}, \citenamefont
		{{Kuroki}}, \citenamefont {{Eisaki}}, \citenamefont {{Uchida}},\ and\
		\citenamefont {{Aoki}}}]{Yamazaki2020}%
	\BibitemOpen
	\bibfield  {author} {\bibinfo {author} {\bibfnamefont {K.}~\bibnamefont
			{{Yamazaki}}}, \bibinfo {author} {\bibfnamefont {M.}~\bibnamefont {{Ochi}}},
		\bibinfo {author} {\bibfnamefont {D.}~\bibnamefont {{Ogura}}}, \bibinfo
		{author} {\bibfnamefont {K.}~\bibnamefont {{Kuroki}}}, \bibinfo {author}
		{\bibfnamefont {H.}~\bibnamefont {{Eisaki}}}, \bibinfo {author}
		{\bibfnamefont {S.}~\bibnamefont {{Uchida}}},\ and\ \bibinfo {author}
		{\bibfnamefont {H.}~\bibnamefont {{Aoki}}},\ }\bibfield  {title} {\bibinfo
		{title} {{Superconducting mechanism for the cuprate Ba$_{2}$Cu O$_{3
					+{\ensuremath{\delta}}}$ based on a multiorbital Lieb lattice model}},\
	}\href {https://doi.org/10.1103/PhysRevResearch.2.033356} {\bibfield
		{journal} {\bibinfo  {journal} {Physical Review Research}\ }\textbf {\bibinfo
			{volume} {2}},\ \bibinfo {eid} {033356} (\bibinfo {year} {2020})}\BibitemShut
	{NoStop}%
	\bibitem [{\citenamefont {{Nam}}\ and\ \citenamefont
		{{Ardavan}}(2020)}]{Nam2020}%
	\BibitemOpen
	\bibfield  {author} {\bibinfo {author} {\bibfnamefont {M.-S.}\ \bibnamefont
			{{Nam}}}\ and\ \bibinfo {author} {\bibfnamefont {A.}~\bibnamefont
			{{Ardavan}}},\ }\bibfield  {title} {\bibinfo {title} {{Universal limiting
				transition temperature for the high $T_\mathrm{c}$ superconductors}},\
	}\href@noop {} {\bibfield  {journal} {\bibinfo  {journal} {arXiv e-prints}\
			,\ \bibinfo {eid} {arXiv:2010.00572}} (\bibinfo {year} {2020})},\ \Eprint
	{https://arxiv.org/abs/2010.00572} {arXiv:2010.00572 [cond-mat.supr-con]}
	\BibitemShut {NoStop}%
	\bibitem [{\citenamefont {{Tan}}\ \emph {et~al.}(2021)\citenamefont {{Tan}},
		\citenamefont {{Liu}}, \citenamefont {{Mou}},\ and\ \citenamefont
		{{Feng}}}]{Tan2021}%
	\BibitemOpen
	\bibfield  {author} {\bibinfo {author} {\bibfnamefont {S.}~\bibnamefont
			{{Tan}}}, \bibinfo {author} {\bibfnamefont {Y.}~\bibnamefont {{Liu}}},
		\bibinfo {author} {\bibfnamefont {Y.}~\bibnamefont {{Mou}}},\ and\ \bibinfo
		{author} {\bibfnamefont {S.}~\bibnamefont {{Feng}}},\ }\bibfield  {title}
	{\bibinfo {title} {{Anisotropic dressing of electrons in electron-doped
				cuprate superconductors}},\ }\href
	{https://doi.org/10.1103/PhysRevB.103.014503} {\bibfield  {journal} {\bibinfo
			{journal} {\prb}\ }\textbf {\bibinfo {volume} {103}},\ \bibinfo {eid}
		{014503} (\bibinfo {year} {2021})}\BibitemShut {NoStop}%
	\bibitem [{\citenamefont {Kyung}\ \emph {et~al.}(2009)\citenamefont {Kyung},
		\citenamefont {S\'en\'echal},\ and\ \citenamefont {Tremblay}}]{Kyung2009}%
	\BibitemOpen
	\bibfield  {author} {\bibinfo {author} {\bibfnamefont {B.}~\bibnamefont
			{Kyung}}, \bibinfo {author} {\bibfnamefont {D.}~\bibnamefont
			{S\'en\'echal}},\ and\ \bibinfo {author} {\bibfnamefont {A.-M.~S.}\
			\bibnamefont {Tremblay}},\ }\bibfield  {title} {\bibinfo {title} {Pairing
			dynamics in strongly correlated superconductivity},\ }\href
	{https://doi.org/10.1103/PhysRevB.80.205109} {\bibfield  {journal} {\bibinfo
			{journal} {Phys. Rev. B}\ }\textbf {\bibinfo {volume} {80}},\ \bibinfo
		{pages} {205109} (\bibinfo {year} {2009})}\BibitemShut {NoStop}%
	\bibitem [{\citenamefont {{Capone}}\ \emph {et~al.}(2002)\citenamefont
		{{Capone}}, \citenamefont {{Fabrizio}}, \citenamefont {{Castellani}},\ and\
		\citenamefont {{Tosatti}}}]{Capone2012}%
	\BibitemOpen
	\bibfield  {author} {\bibinfo {author} {\bibfnamefont {M.}~\bibnamefont
			{{Capone}}}, \bibinfo {author} {\bibfnamefont {M.}~\bibnamefont
			{{Fabrizio}}}, \bibinfo {author} {\bibfnamefont {C.}~\bibnamefont
			{{Castellani}}},\ and\ \bibinfo {author} {\bibfnamefont {E.}~\bibnamefont
			{{Tosatti}}},\ }\bibfield  {title} {\bibinfo {title} {{Strongly Correlated
				Superconductivity}},\ }\href {https://doi.org/10.1126/science.1071122}
	{\bibfield  {journal} {\bibinfo  {journal} {Science}\ }\textbf {\bibinfo
			{volume} {296}},\ \bibinfo {pages} {2364} (\bibinfo {year}
		{2002})}\BibitemShut {NoStop}%
	\bibitem [{\citenamefont {Cao}\ \emph {et~al.}(2018)\citenamefont {Cao},
		\citenamefont {Fatemi}, \citenamefont {Fang}, \citenamefont {Watanabe},
		\citenamefont {Taniguchi}, \citenamefont {Kaxiras},\ and\ \citenamefont
		{Jarillo-Herrero}}]{Cao2018}%
	\BibitemOpen
	\bibfield  {author} {\bibinfo {author} {\bibfnamefont {Y.}~\bibnamefont
			{Cao}}, \bibinfo {author} {\bibfnamefont {V.}~\bibnamefont {Fatemi}},
		\bibinfo {author} {\bibfnamefont {S.}~\bibnamefont {Fang}}, \bibinfo {author}
		{\bibfnamefont {K.}~\bibnamefont {Watanabe}}, \bibinfo {author}
		{\bibfnamefont {T.}~\bibnamefont {Taniguchi}}, \bibinfo {author}
		{\bibfnamefont {E.}~\bibnamefont {Kaxiras}},\ and\ \bibinfo {author}
		{\bibfnamefont {P.}~\bibnamefont {Jarillo-Herrero}},\ }\bibfield  {title}
	{\bibinfo {title} {Unconventional superconductivity in magic-angle graphene
			superlattices},\ }\href {https://doi.org/10.1038/nature26160} {\bibfield
		{journal} {\bibinfo  {journal} {Nature}\ }\textbf {\bibinfo {volume} {556}},\
		\bibinfo {pages} {43} (\bibinfo {year} {2018})}\BibitemShut {NoStop}%
	\bibitem [{\citenamefont {{Yankowitz}}\ \emph {et~al.}(2019)\citenamefont
		{{Yankowitz}}, \citenamefont {{Chen}}, \citenamefont {{Polshyn}},
		\citenamefont {{Zhang}}, \citenamefont {{Watanabe}}, \citenamefont
		{{Taniguchi}}, \citenamefont {{Graf}}, \citenamefont {{Young}},\ and\
		\citenamefont {{Dean}}}]{Yankowitz2019}%
	\BibitemOpen
	\bibfield  {author} {\bibinfo {author} {\bibfnamefont {M.}~\bibnamefont
			{{Yankowitz}}}, \bibinfo {author} {\bibfnamefont {S.}~\bibnamefont {{Chen}}},
		\bibinfo {author} {\bibfnamefont {H.}~\bibnamefont {{Polshyn}}}, \bibinfo
		{author} {\bibfnamefont {Y.}~\bibnamefont {{Zhang}}}, \bibinfo {author}
		{\bibfnamefont {K.}~\bibnamefont {{Watanabe}}}, \bibinfo {author}
		{\bibfnamefont {T.}~\bibnamefont {{Taniguchi}}}, \bibinfo {author}
		{\bibfnamefont {D.}~\bibnamefont {{Graf}}}, \bibinfo {author} {\bibfnamefont
			{A.~F.}\ \bibnamefont {{Young}}},\ and\ \bibinfo {author} {\bibfnamefont
			{C.~R.}\ \bibnamefont {{Dean}}},\ }\bibfield  {title} {\bibinfo {title}
		{{Tuning superconductivity in twisted bilayer graphene}},\ }\href
	{https://doi.org/10.1126/science.aav1910} {\bibfield  {journal} {\bibinfo
			{journal} {Science}\ }\textbf {\bibinfo {volume} {363}},\ \bibinfo {pages}
		{1059} (\bibinfo {year} {2019})}\BibitemShut {NoStop}%
	\bibitem [{\citenamefont {Kerelsky}\ \emph {et~al.}(2019)\citenamefont
		{Kerelsky}, \citenamefont {McGilly},\ and\ \citenamefont
		{Kennes}}]{Kerelsky2019}%
	\BibitemOpen
	\bibfield  {author} {\bibinfo {author} {\bibfnamefont {A.}~\bibnamefont
			{Kerelsky}}, \bibinfo {author} {\bibfnamefont {L.}~\bibnamefont {McGilly}},\
		and\ \bibinfo {author} {\bibfnamefont {D.~e.~a.}\ \bibnamefont {Kennes}},\
	}\bibfield  {title} {\bibinfo {title} {{Maximized electron interactions at
				the magic angle in twisted bilayer graphene}},\ }\href
	{https://doi.org/10.1038/s41586-019-1431-9} {\bibfield  {journal} {\bibinfo
			{journal} {Nature}\ }\textbf {\bibinfo {volume} {572}},\ \bibinfo {pages}
		{95} (\bibinfo {year} {2019})}\BibitemShut {NoStop}%
	\bibitem [{\citenamefont {Bistritzer}\ and\ \citenamefont
		{MacDonald}(2011)}]{MacDonald2011}%
	\BibitemOpen
	\bibfield  {author} {\bibinfo {author} {\bibfnamefont {R.}~\bibnamefont
			{Bistritzer}}\ and\ \bibinfo {author} {\bibfnamefont {A.~H.}\ \bibnamefont
			{MacDonald}},\ }\bibfield  {title} {\bibinfo {title} {Moir{\'e} bands in
			twisted double-layer graphene},\ }\href
	{https://doi.org/10.1073/pnas.1108174108} {\bibfield  {journal} {\bibinfo
			{journal} {Proceedings of the National Academy of Sciences}\ }\textbf
		{\bibinfo {volume} {108}},\ \bibinfo {pages} {12233} (\bibinfo {year}
		{2011})}\BibitemShut {NoStop}%
	\bibitem [{\citenamefont {dos Santos}\ \emph {et~al.}(2007)\citenamefont {dos
			Santos}, \citenamefont {Peres},\ and\ \citenamefont {Neto}}]{Santos2007}%
	\BibitemOpen
	\bibfield  {author} {\bibinfo {author} {\bibfnamefont {J.~M. B.~L.}\
			\bibnamefont {dos Santos}}, \bibinfo {author} {\bibfnamefont {N.~M.~R.}\
			\bibnamefont {Peres}},\ and\ \bibinfo {author} {\bibfnamefont {A.~H.~C.}\
			\bibnamefont {Neto}},\ }\bibfield  {title} {\bibinfo {title} {Graphene
			bilayer with a twist: Electronic structure},\ }\href
	{https://doi.org/https://doi.org/10.1103/PhysRevLett.99.256802} {\bibfield
		{journal} {\bibinfo  {journal} {Phys. Rev. Lett.}\ }\textbf {\bibinfo
			{volume} {99}},\ \bibinfo {pages} {256802} (\bibinfo {year}
		{2007})}\BibitemShut {NoStop}%
	\bibitem [{\citenamefont {Tarnopolsky}\ \emph {et~al.}(2019)\citenamefont
		{Tarnopolsky}, \citenamefont {Kruchkov},\ and\ \citenamefont
		{Vishwanath}}]{Tarnpolsky2019}%
	\BibitemOpen
	\bibfield  {author} {\bibinfo {author} {\bibfnamefont {G.}~\bibnamefont
			{Tarnopolsky}}, \bibinfo {author} {\bibfnamefont {A.~J.}\ \bibnamefont
			{Kruchkov}},\ and\ \bibinfo {author} {\bibfnamefont {A.}~\bibnamefont
			{Vishwanath}},\ }\bibfield  {title} {\bibinfo {title} {Origin of magic angles
			in twisted bilayer graphene},\ }\href
	{https://doi.org/10.1103/PhysRevLett.122.106405} {\bibfield  {journal}
		{\bibinfo  {journal} {Phys. Rev. Lett.}\ }\textbf {\bibinfo {volume} {122}},\
		\bibinfo {pages} {106405} (\bibinfo {year} {2019})}\BibitemShut {NoStop}%
	\bibitem [{\citenamefont {{Wang}}\ \emph
		{et~al.}(2020{\natexlab{b}})\citenamefont {{Wang}}, \citenamefont {{Zheng}},
		\citenamefont {{Millis}},\ and\ \citenamefont {{Cano}}}]{Wang2020}%
	\BibitemOpen
	\bibfield  {author} {\bibinfo {author} {\bibfnamefont {J.}~\bibnamefont
			{{Wang}}}, \bibinfo {author} {\bibfnamefont {Y.}~\bibnamefont {{Zheng}}},
		\bibinfo {author} {\bibfnamefont {A.~J.}\ \bibnamefont {{Millis}}},\ and\
		\bibinfo {author} {\bibfnamefont {J.}~\bibnamefont {{Cano}}},\ }\bibfield
	{title} {\bibinfo {title} {{Chiral Approximation to Twisted Bilayer Graphene:
				Exact Intra-Valley Inversion Symmetry, Nodal Structure and Implications for
				Higher Magic Angles}},\ }\href@noop {} {\bibfield  {journal} {\bibinfo
			{journal} {arXiv e-prints}\ ,\ \bibinfo {eid} {arXiv:2010.03589}} (\bibinfo
		{year} {2020}{\natexlab{b}})},\ \Eprint {https://arxiv.org/abs/2010.03589}
	{arXiv:2010.03589 [cond-mat.mes-hall]} \BibitemShut {NoStop}%
	\bibitem [{\citenamefont {Popov}\ and\ \citenamefont
		{Milekhin}(2020)}]{Popov2020}%
	\BibitemOpen
	\bibfield  {author} {\bibinfo {author} {\bibfnamefont {F.~K.}\ \bibnamefont
			{Popov}}\ and\ \bibinfo {author} {\bibfnamefont {A.}~\bibnamefont
			{Milekhin}},\ }\bibfield  {title} {\bibinfo {title} {Hidden wave function of
			twisted bilayer graphene: Flat band as a landau level},\ }\href@noop {}
	{\bibfield  {journal} {\bibinfo  {journal} {arXiv e-prints}\ ,\ \bibinfo
			{eid} {arXiv:2010.02915}} (\bibinfo {year} {2020})},\ \Eprint
	{https://arxiv.org/abs/2010.02915v1} {arXiv:2010.02915v1 [cond-mat.mes-hall]}
	\BibitemShut {NoStop}%
	\bibitem [{\citenamefont {{Hejazi}}\ \emph {et~al.}(2019)\citenamefont
		{{Hejazi}}, \citenamefont {{Liu}},\ and\ \citenamefont
		{{Balents}}}]{Hejazi2019}%
	\BibitemOpen
	\bibfield  {author} {\bibinfo {author} {\bibfnamefont {K.}~\bibnamefont
			{{Hejazi}}}, \bibinfo {author} {\bibfnamefont {C.}~\bibnamefont {{Liu}}},\
		and\ \bibinfo {author} {\bibfnamefont {L.}~\bibnamefont {{Balents}}},\
	}\bibfield  {title} {\bibinfo {title} {{Landau levels in twisted bilayer
				graphene and semiclassical orbits}},\ }\href
	{https://doi.org/10.1103/PhysRevB.100.035115} {\bibfield  {journal} {\bibinfo
			{journal} {\prb}\ }\textbf {\bibinfo {volume} {100}},\ \bibinfo {eid}
		{035115} (\bibinfo {year} {2019})}\BibitemShut {NoStop}%
	\bibitem [{\citenamefont {{Uri}}\ \emph {et~al.}(2020)\citenamefont {{Uri}},
		\citenamefont {{Grover}}, \citenamefont {{Cao}}, \citenamefont {{Crosse}},
		\citenamefont {{Bagani}}, \citenamefont {{Rodan-Legrain}}, \citenamefont
		{{Myasoedov}}, \citenamefont {{Watanabe}}, \citenamefont {{Taniguchi}},
		\citenamefont {{Moon}}, \citenamefont {{Koshino}}, \citenamefont
		{{Jarillo-Herrero}},\ and\ \citenamefont {{Zeldov}}}]{Uri2020}%
	\BibitemOpen
	\bibfield  {author} {\bibinfo {author} {\bibfnamefont {A.}~\bibnamefont
			{{Uri}}}, \bibinfo {author} {\bibfnamefont {S.}~\bibnamefont {{Grover}}},
		\bibinfo {author} {\bibfnamefont {Y.}~\bibnamefont {{Cao}}}, \bibinfo
		{author} {\bibfnamefont {J.~{\^A}.~A.}\ \bibnamefont {{Crosse}}}, \bibinfo
		{author} {\bibfnamefont {K.}~\bibnamefont {{Bagani}}}, \bibinfo {author}
		{\bibfnamefont {D.}~\bibnamefont {{Rodan-Legrain}}}, \bibinfo {author}
		{\bibfnamefont {Y.}~\bibnamefont {{Myasoedov}}}, \bibinfo {author}
		{\bibfnamefont {K.}~\bibnamefont {{Watanabe}}}, \bibinfo {author}
		{\bibfnamefont {T.}~\bibnamefont {{Taniguchi}}}, \bibinfo {author}
		{\bibfnamefont {P.}~\bibnamefont {{Moon}}}, \bibinfo {author} {\bibfnamefont
			{M.}~\bibnamefont {{Koshino}}}, \bibinfo {author} {\bibfnamefont
			{P.}~\bibnamefont {{Jarillo-Herrero}}},\ and\ \bibinfo {author}
		{\bibfnamefont {E.}~\bibnamefont {{Zeldov}}},\ }\bibfield  {title} {\bibinfo
		{title} {{Mapping the twist-angle disorder and Landau levels in magic-angle
				graphene}},\ }\href {https://doi.org/10.1038/s41586-020-2255-3} {\bibfield
		{journal} {\bibinfo  {journal} {\nat}\ }\textbf {\bibinfo {volume} {581}},\
		\bibinfo {pages} {47} (\bibinfo {year} {2020})}\BibitemShut {NoStop}%
	\bibitem [{\citenamefont {{Qi}}\ and\ \citenamefont {{Zhang}}(2011)}]{Qi2011}%
	\BibitemOpen
	\bibfield  {author} {\bibinfo {author} {\bibfnamefont {X.-L.}\ \bibnamefont
			{{Qi}}}\ and\ \bibinfo {author} {\bibfnamefont {S.-C.}\ \bibnamefont
			{{Zhang}}},\ }\bibfield  {title} {\bibinfo {title} {{Topological insulators
				and superconductors}},\ }\href {https://doi.org/10.1103/RevModPhys.83.1057}
	{\bibfield  {journal} {\bibinfo  {journal} {Reviews of Modern Physics}\
		}\textbf {\bibinfo {volume} {83}},\ \bibinfo {pages} {1057} (\bibinfo {year}
		{2011})}\BibitemShut {NoStop}%
	\bibitem [{\citenamefont {{Zou}}\ \emph {et~al.}(2018)\citenamefont {{Zou}},
		\citenamefont {{Po}}, \citenamefont {{Vishwanath}},\ and\ \citenamefont
		{{Senthil}}}]{Zou2018}%
	\BibitemOpen
	\bibfield  {author} {\bibinfo {author} {\bibfnamefont {L.}~\bibnamefont
			{{Zou}}}, \bibinfo {author} {\bibfnamefont {H.~C.}\ \bibnamefont {{Po}}},
		\bibinfo {author} {\bibfnamefont {A.}~\bibnamefont {{Vishwanath}}},\ and\
		\bibinfo {author} {\bibfnamefont {T.}~\bibnamefont {{Senthil}}},\ }\bibfield
	{title} {\bibinfo {title} {{Band structure of twisted bilayer graphene:
				Emergent symmetries, commensurate approximants, and Wannier obstructions}},\
	}\bibfield  {journal} {\bibinfo  {journal} {\prb}\ }\textbf {\bibinfo
		{volume} {98}},\ \href {https://doi.org/10.1103/PhysRevB.98.085435}
	{10.1103/PhysRevB.98.085435} (\bibinfo {year} {2018})\BibitemShut {NoStop}%
	\bibitem [{\citenamefont {Xu}\ and\ \citenamefont {Balents}(2018)}]{Xu2018}%
	\BibitemOpen
	\bibfield  {author} {\bibinfo {author} {\bibfnamefont {C.}~\bibnamefont
			{Xu}}\ and\ \bibinfo {author} {\bibfnamefont {L.}~\bibnamefont {Balents}},\
	}\bibfield  {title} {\bibinfo {title} {Topological superconductivity in
			twisted multilayer graphene},\ }\href
	{https://doi.org/10.1103/PhysRevLett.121.087001} {\bibfield  {journal}
		{\bibinfo  {journal} {Phys. Rev. Lett.}\ }\textbf {\bibinfo {volume} {121}},\
		\bibinfo {pages} {087001} (\bibinfo {year} {2018})}\BibitemShut {NoStop}%
	\bibitem [{\citenamefont {{Fidrysiak}}\ \emph {et~al.}(2018)\citenamefont
		{{Fidrysiak}}, \citenamefont {{Zegrodnik}},\ and\ \citenamefont
		{{Spa{\l}ek}}}]{Fidrysiak2018}%
	\BibitemOpen
	\bibfield  {author} {\bibinfo {author} {\bibfnamefont {M.}~\bibnamefont
			{{Fidrysiak}}}, \bibinfo {author} {\bibfnamefont {M.}~\bibnamefont
			{{Zegrodnik}}},\ and\ \bibinfo {author} {\bibfnamefont {J.}~\bibnamefont
			{{Spa{\l}ek}}},\ }\bibfield  {title} {\bibinfo {title} {{Unconventional
				topological superconductivity and phase diagram for an effective two-orbital
				model as applied to twisted bilayer graphene}},\ }\href
	{https://doi.org/10.1103/PhysRevB.98.085436} {\bibfield  {journal} {\bibinfo
			{journal} {\prb}\ }\textbf {\bibinfo {volume} {98}},\ \bibinfo {eid} {085436}
		(\bibinfo {year} {2018})}\BibitemShut {NoStop}%
	\bibitem [{\citenamefont {{You}}\ and\ \citenamefont
		{{Vishwanath}}(2019)}]{You2019}%
	\BibitemOpen
	\bibfield  {author} {\bibinfo {author} {\bibfnamefont {Y.-Z.}\ \bibnamefont
			{{You}}}\ and\ \bibinfo {author} {\bibfnamefont {A.}~\bibnamefont
			{{Vishwanath}}},\ }\bibfield  {title} {\bibinfo {title} {{Superconductivity
				from valley fluctuations and approximate SO(4) symmetry in a weak coupling
				theory of twisted bilayer graphene}},\ }\href
	{https://doi.org/10.1038/s41535-019-0153-4} {\bibfield  {journal} {\bibinfo
			{journal} {npj Quantum Materials}\ }\textbf {\bibinfo {volume} {4}},\
		\bibinfo {eid} {16} (\bibinfo {year} {2019})}\BibitemShut {NoStop}%
	\bibitem [{\citenamefont {{Khalaf}}\ \emph {et~al.}(2020)\citenamefont
		{{Khalaf}}, \citenamefont {{Chatterjee}}, \citenamefont {{Bultinck}},
		\citenamefont {{Zaletel}},\ and\ \citenamefont {{Vishwanath}}}]{Khalaf2020}%
	\BibitemOpen
	\bibfield  {author} {\bibinfo {author} {\bibfnamefont {E.}~\bibnamefont
			{{Khalaf}}}, \bibinfo {author} {\bibfnamefont {S.}~\bibnamefont
			{{Chatterjee}}}, \bibinfo {author} {\bibfnamefont {N.}~\bibnamefont
			{{Bultinck}}}, \bibinfo {author} {\bibfnamefont {M.~P.}\ \bibnamefont
			{{Zaletel}}},\ and\ \bibinfo {author} {\bibfnamefont {A.}~\bibnamefont
			{{Vishwanath}}},\ }\bibfield  {title} {\bibinfo {title} {{Charged Skyrmions
				and Topological Origin of Superconductivity in Magic Angle Graphene}},\
	}\href@noop {} {\bibfield  {journal} {\bibinfo  {journal} {arXiv e-prints}\
			,\ \bibinfo {eid} {arXiv:2004.00638}} (\bibinfo {year} {2020})},\ \Eprint
	{https://arxiv.org/abs/2004.00638} {arXiv:2004.00638 [cond-mat.str-el]}
	\BibitemShut {NoStop}%
	\bibitem [{\citenamefont {Rodriguez-Vega}\ \emph {et~al.}(2020)\citenamefont
		{Rodriguez-Vega}, \citenamefont {Vogl},\ and\ \citenamefont
		{Fiete}}]{Vogl2020}%
	\BibitemOpen
	\bibfield  {author} {\bibinfo {author} {\bibfnamefont {M.}~\bibnamefont
			{Rodriguez-Vega}}, \bibinfo {author} {\bibfnamefont {M.}~\bibnamefont
			{Vogl}},\ and\ \bibinfo {author} {\bibfnamefont {G.~A.}\ \bibnamefont
			{Fiete}},\ }\bibfield  {title} {\bibinfo {title} {Floquet engineering of
			twisted double bilayer graphene},\ }\href
	{https://doi.org/10.1103/PhysRevResearch.2.033494} {\bibfield  {journal}
		{\bibinfo  {journal} {Phys. Rev. Research}\ }\textbf {\bibinfo {volume}
			{2}},\ \bibinfo {pages} {033494} (\bibinfo {year} {2020})}\BibitemShut
	{NoStop}%
	\bibitem [{\citenamefont {Naumis}(2007)}]{Naumis2007}%
	\BibitemOpen
	\bibfield  {author} {\bibinfo {author} {\bibfnamefont {G.~G.}\ \bibnamefont
			{Naumis}},\ }\bibfield  {title} {\bibinfo {title} {Internal mobility edge in
			doped graphene: Frustration in a renormalized lattice},\ }\href
	{https://doi.org/10.1103/PhysRevB.76.153403} {\bibfield  {journal} {\bibinfo
			{journal} {Phys. Rev. B}\ }\textbf {\bibinfo {volume} {76}},\ \bibinfo
		{pages} {153403} (\bibinfo {year} {2007})}\BibitemShut {NoStop}%
	\bibitem [{\citenamefont {Barrios-Vargas}\ and\ \citenamefont
		{Naumis}(2011)}]{Barrios_Vargas_2011}%
	\BibitemOpen
	\bibfield  {author} {\bibinfo {author} {\bibfnamefont {J.~E.}\ \bibnamefont
			{Barrios-Vargas}}\ and\ \bibinfo {author} {\bibfnamefont {G.~G.}\
			\bibnamefont {Naumis}},\ }\bibfield  {title} {\bibinfo {title} {Doped
			graphene: the interplay between localization and frustration due to the
			underlying triangular symmetry},\ }\href
	{https://doi.org/10.1088/0953-8984/23/37/375501} {\bibfield  {journal}
		{\bibinfo  {journal} {Journal of Physics: Condensed Matter}\ }\textbf
		{\bibinfo {volume} {23}},\ \bibinfo {pages} {375501} (\bibinfo {year}
		{2011})}\BibitemShut {NoStop}%
	\bibitem [{\citenamefont {Fumiko~Yonezawa}(1981)}]{Cohen1981}%
	\BibitemOpen
	\bibfield  {author} {\bibinfo {author} {\bibfnamefont {M.~C.}\ \bibnamefont
			{Fumiko~Yonezawa}},\ }\bibfield  {title} {\bibinfo {title} {Theory of
			electronic properties of amorphous semiconductors},\ }in\ \href@noop {}
	{\emph {\bibinfo {booktitle} {Fundamental Physics of Amorphous
				Semiconductors: Proceedings of the Kyoto Summer Institute Kyoto, Japan,
				September 8—11, 1980}}},\ \bibinfo {series} {Springer Series in Solid-State
		Sciences}, Vol.~\bibinfo {volume} {46},\ \bibinfo {editor} {edited by\
		\bibinfo {editor} {\bibfnamefont {F.~Y.}\ \bibnamefont {Fumiko~Yonezawa}}}\
	(\bibinfo  {publisher} {Springer-Verlag Berlin Heidelberg},\ \bibinfo {year}
	{1981})\ \bibinfo {edition} {1st}\ ed.,\ pp.\ \bibinfo {pages}
	{119--144}\BibitemShut {NoStop}%
	\bibitem [{\citenamefont {Cohen}(1983)}]{Cohen1983}%
	\BibitemOpen
	\bibfield  {author} {\bibinfo {author} {\bibfnamefont {M.}~\bibnamefont
			{Cohen}},\ }\bibfield  {title} {\bibinfo {title} {Topology, geometry,
			elementary excitations and physical properties of disordered materials},\
	}in\ \href@noop {} {\emph {\bibinfo {booktitle} {Topological Disorder in
				Condensed Matter}}},\ \bibinfo {series} {Springer Series in Solid-State
		Sciences}, Vol.~\bibinfo {volume} {25},\ \bibinfo {editor} {edited by\
		\bibinfo {editor} {\bibfnamefont {F.~Y.}\ \bibnamefont {Hellmut~Fritzsche}}}\
	(\bibinfo  {publisher} {Springer-Verlag Berlin Heidelberg},\ \bibinfo {year}
	{1983})\ \bibinfo {edition} {1st}\ ed.,\ pp.\ \bibinfo {pages}
	{122--141}\BibitemShut {NoStop}%
	\bibitem [{\citenamefont {Kohmoto}\ and\ \citenamefont
		{Sutherland}(1986)}]{Kohmoto1986}%
	\BibitemOpen
	\bibfield  {author} {\bibinfo {author} {\bibfnamefont {M.}~\bibnamefont
			{Kohmoto}}\ and\ \bibinfo {author} {\bibfnamefont {B.}~\bibnamefont
			{Sutherland}},\ }\bibfield  {title} {\bibinfo {title} {Electronic states on a
			penrose lattice},\ }\href {https://doi.org/10.1103/PhysRevLett.56.2740}
	{\bibfield  {journal} {\bibinfo  {journal} {Phys. Rev. Lett.}\ }\textbf
		{\bibinfo {volume} {56}},\ \bibinfo {pages} {2740} (\bibinfo {year}
		{1986})}\BibitemShut {NoStop}%
	\bibitem [{\citenamefont {Arai}\ \emph {et~al.}(1988)\citenamefont {Arai},
		\citenamefont {Tokihiro}, \citenamefont {Fujiwara},\ and\ \citenamefont
		{Kohmoto}}]{Arai1988}%
	\BibitemOpen
	\bibfield  {author} {\bibinfo {author} {\bibfnamefont {M.}~\bibnamefont
			{Arai}}, \bibinfo {author} {\bibfnamefont {T.}~\bibnamefont {Tokihiro}},
		\bibinfo {author} {\bibfnamefont {T.}~\bibnamefont {Fujiwara}},\ and\
		\bibinfo {author} {\bibfnamefont {M.}~\bibnamefont {Kohmoto}},\ }\bibfield
	{title} {\bibinfo {title} {Strictly localized states on a two-dimensional
			penrose lattice},\ }\href {https://doi.org/10.1103/PhysRevB.38.1621}
	{\bibfield  {journal} {\bibinfo  {journal} {Phys. Rev. B}\ }\textbf {\bibinfo
			{volume} {38}},\ \bibinfo {pages} {1621} (\bibinfo {year}
		{1988})}\BibitemShut {NoStop}%
	\bibitem [{\citenamefont {Naumis}\ \emph {et~al.}(1994)\citenamefont {Naumis},
		\citenamefont {Barrio},\ and\ \citenamefont {Wang}}]{Naumis1994}%
	\BibitemOpen
	\bibfield  {author} {\bibinfo {author} {\bibfnamefont {G.~G.}\ \bibnamefont
			{Naumis}}, \bibinfo {author} {\bibfnamefont {R.~A.}\ \bibnamefont {Barrio}},\
		and\ \bibinfo {author} {\bibfnamefont {C.}~\bibnamefont {Wang}},\ }\bibfield
	{title} {\bibinfo {title} {Effects of frustration and localization of states
			in the penrose lattice},\ }\href {https://doi.org/10.1103/PhysRevB.50.9834}
	{\bibfield  {journal} {\bibinfo  {journal} {Phys. Rev. B}\ }\textbf {\bibinfo
			{volume} {50}},\ \bibinfo {pages} {9834} (\bibinfo {year}
		{1994})}\BibitemShut {NoStop}%
	\bibitem [{\citenamefont {Kirkpatrick}\ and\ \citenamefont
		{Eggarter}(1972)}]{Eggarter1972}%
	\BibitemOpen
	\bibfield  {author} {\bibinfo {author} {\bibfnamefont {S.}~\bibnamefont
			{Kirkpatrick}}\ and\ \bibinfo {author} {\bibfnamefont {T.~P.}\ \bibnamefont
			{Eggarter}},\ }\bibfield  {title} {\bibinfo {title} {Localized states of a
			binary alloy},\ }\href {https://doi.org/10.1103/PhysRevB.6.3598} {\bibfield
		{journal} {\bibinfo  {journal} {Phys. Rev. B}\ }\textbf {\bibinfo {volume}
			{6}},\ \bibinfo {pages} {3598} (\bibinfo {year} {1972})}\BibitemShut
	{NoStop}%
	\bibitem [{\citenamefont {Naumis}\ \emph {et~al.}(2002)\citenamefont {Naumis},
		\citenamefont {Wang},\ and\ \citenamefont {Barrio}}]{Naumis2002}%
	\BibitemOpen
	\bibfield  {author} {\bibinfo {author} {\bibfnamefont {G.~G.}\ \bibnamefont
			{Naumis}}, \bibinfo {author} {\bibfnamefont {C.}~\bibnamefont {Wang}},\ and\
		\bibinfo {author} {\bibfnamefont {R.~A.}\ \bibnamefont {Barrio}},\ }\bibfield
	{title} {\bibinfo {title} {Frustration effects on the electronic density of
			states of a random binary alloy},\ }\href
	{https://doi.org/10.1103/PhysRevB.65.134203} {\bibfield  {journal} {\bibinfo
			{journal} {Phys. Rev. B}\ }\textbf {\bibinfo {volume} {65}},\ \bibinfo
		{pages} {134203} (\bibinfo {year} {2002})}\BibitemShut {NoStop}%
	\bibitem [{\citenamefont {Barrios-Vargas}\ and\ \citenamefont
		{Naumis}(2013)}]{Barrios2013}%
	\BibitemOpen
	\bibfield  {author} {\bibinfo {author} {\bibfnamefont {J.}~\bibnamefont
			{Barrios-Vargas}}\ and\ \bibinfo {author} {\bibfnamefont {G.~G.}\
			\bibnamefont {Naumis}},\ }\bibfield  {title} {\bibinfo {title} {Pseudo-gap
			opening and dirac point confined states in doped graphene},\ }\href
	{https://doi.org/https://doi.org/10.1016/j.ssc.2013.03.006} {\bibfield
		{journal} {\bibinfo  {journal} {Solid State Communications}\ }\textbf
		{\bibinfo {volume} {162}},\ \bibinfo {pages} {23} (\bibinfo {year}
		{2013})}\BibitemShut {NoStop}%
	\bibitem [{\citenamefont {Kane}\ and\ \citenamefont
		{Lubensky}(2014)}]{Kane2014}%
	\BibitemOpen
	\bibfield  {author} {\bibinfo {author} {\bibfnamefont {C.~L.}\ \bibnamefont
			{Kane}}\ and\ \bibinfo {author} {\bibfnamefont {T.~C.}\ \bibnamefont
			{Lubensky}},\ }\bibfield  {title} {\bibinfo {title} {Topological boundary
			modes in isostatic lattices},\ }\href {https://doi.org/10.1038/nphys2835}
	{\bibfield  {journal} {\bibinfo  {journal} {Nature Physics}\ }\textbf
		{\bibinfo {volume} {10}},\ \bibinfo {pages} {39} (\bibinfo {year}
		{2014})}\BibitemShut {NoStop}%
	\bibitem [{\citenamefont {Phillips}(1979)}]{Phillips}%
	\BibitemOpen
	\bibfield  {author} {\bibinfo {author} {\bibfnamefont {J.~C.}\ \bibnamefont
			{Phillips}},\ }\bibfield  {title} {\bibinfo {title} {Topology of covalent
			non-crystalline solids i: Short-range order in chalcogenide alloys},\ }\href
	{https://doi.org/https://doi.org/10.1016/0022-3093(79)90033-4} {\bibfield
		{journal} {\bibinfo  {journal} {Journal of Non-Crystalline Solids}\ }\textbf
		{\bibinfo {volume} {34}},\ \bibinfo {pages} {153} (\bibinfo {year}
		{1979})}\BibitemShut {NoStop}%
	\bibitem [{\citenamefont {Huerta}\ and\ \citenamefont
		{Naumis}(2002)}]{Huerta2002}%
	\BibitemOpen
	\bibfield  {author} {\bibinfo {author} {\bibfnamefont {A.}~\bibnamefont
			{Huerta}}\ and\ \bibinfo {author} {\bibfnamefont {G.~G.}\ \bibnamefont
			{Naumis}},\ }\bibfield  {title} {\bibinfo {title} {Evidence of a glass
			transition induced by rigidity self-organization in a network-forming
			fluid},\ }\href {https://doi.org/10.1103/PhysRevB.66.184204} {\bibfield
		{journal} {\bibinfo  {journal} {Phys. Rev. B}\ }\textbf {\bibinfo {volume}
			{66}},\ \bibinfo {pages} {184204} (\bibinfo {year} {2002})}\BibitemShut
	{NoStop}%
	\bibitem [{\citenamefont {Flores-Ruiz}\ \emph {et~al.}(2010)\citenamefont
		{Flores-Ruiz}, \citenamefont {Naumis},\ and\ \citenamefont
		{Phillips}}]{Flores2010}%
	\BibitemOpen
	\bibfield  {author} {\bibinfo {author} {\bibfnamefont {H.~M.}\ \bibnamefont
			{Flores-Ruiz}}, \bibinfo {author} {\bibfnamefont {G.~G.}\ \bibnamefont
			{Naumis}},\ and\ \bibinfo {author} {\bibfnamefont {J.~C.}\ \bibnamefont
			{Phillips}},\ }\bibfield  {title} {\bibinfo {title} {Heating through the
			glass transition: A rigidity approach to the boson peak},\ }\href
	{https://doi.org/10.1103/PhysRevB.82.214201} {\bibfield  {journal} {\bibinfo
			{journal} {Phys. Rev. B}\ }\textbf {\bibinfo {volume} {82}},\ \bibinfo
		{pages} {214201} (\bibinfo {year} {2010})}\BibitemShut {NoStop}%
	\bibitem [{\citenamefont {San-Jose}\ \emph {et~al.}(2012)\citenamefont
		{San-Jose}, \citenamefont {González},\ and\ \citenamefont
		{Guinea}}]{Guinea2012}%
	\BibitemOpen
	\bibfield  {author} {\bibinfo {author} {\bibfnamefont {P.}~\bibnamefont
			{San-Jose}}, \bibinfo {author} {\bibfnamefont {J.}~\bibnamefont
			{González}},\ and\ \bibinfo {author} {\bibfnamefont {F.}~\bibnamefont
			{Guinea}},\ }\bibfield  {title} {\bibinfo {title} {Non-abelian gauge
			potentials in graphene bilayers},\ }\href
	{https://doi.org/https://doi.org/10.1103/PhysRevLett.108.216802} {\bibfield
		{journal} {\bibinfo  {journal} {Phys. Rev. Lett.}\ }\textbf {\bibinfo
			{volume} {108}},\ \bibinfo {pages} {216802} (\bibinfo {year}
		{2012})}\BibitemShut {NoStop}%
	\bibitem [{\citenamefont {Dasgupta}\ and\ \citenamefont
		{Tchernyshyov}(2020)}]{Dasgupta}%
	\BibitemOpen
	\bibfield  {author} {\bibinfo {author} {\bibfnamefont {S.}~\bibnamefont
			{Dasgupta}}\ and\ \bibinfo {author} {\bibfnamefont {O.}~\bibnamefont
			{Tchernyshyov}},\ }\bibfield  {title} {\bibinfo {title} {Theory of spin waves
			in a hexagonal antiferromagnet},\ }\href
	{https://doi.org/10.1103/PhysRevB.102.144417} {\bibfield  {journal} {\bibinfo
			{journal} {Phys. Rev. B}\ }\textbf {\bibinfo {volume} {102}},\ \bibinfo
		{pages} {144417} (\bibinfo {year} {2020})}\BibitemShut {NoStop}%
\end{thebibliography}


\end{document}